\documentclass[preprint, aps, prmaterials, onecolumn, superscriptaddress, floatfix, altaffilletter, longbibliography, amsmath]{revtex4-2}
\usepackage{graphicx} %
\usepackage{subcaption} %
\usepackage{booktabs} %
\usepackage{makecell} %
\usepackage{multirow} %
\usepackage{xcolor} %
\usepackage{soul} %
\usepackage{verbatim} %
\usepackage{hyperref} %
\usepackage{float} %
\usepackage{placeins} %
\hypersetup{
    colorlinks = true,
    citecolor = blue,
    linkcolor = blue
}

\begin{document}

\title{Resolving Discrepancies in  Calculations of Mechanical Properties of CH\textsubscript{3}NH\textsubscript{3}PbI\textsubscript{3} Perovskites}
\author{Kuntal Talit}
\affiliation{Department of Materials Science and Engineering, University of California, Merced, CA 95343}
\affiliation{Department of Physics, University of California, Merced, CA 95343}
\author{David A. Strubbe}
\email{dstrubbe@ucmerced.edu}
\affiliation{Department of Physics, University of California, Merced, CA 95343}

\begin{abstract}
The mechanical properties of hybrid perovskite materials are important for device flexibility, resistance to fracture, epitaxial growth, surface energetics of quantum dots, and induction or relief of stress in thin films due to thermal expansion and phase changes. These issues are particularly salient for solar cells in space applications. Nonetheless, few studies are available on the mechanical properties of the paradigmatic hybrid perovskite CH$_3$NH$_3$PbI$_3$ (MAPI). Experimental results are only available for the room-temperature tetragonal phase, and there are significant discrepancies among them.  Results from density functional theory (DFT) are available for all three phases but have even larger discrepancies from each other and from experiments. To clarify the disorder in the literature, we have studied the elastic properties of all three phases in detail with DFT calculations. We have examined the effect of different aspects in calculation methodology such as use of energy or stress, the structure, exchange-correlation functionals, Van der Waals corrections, pseudopotentials, $k$-point sampling, and formulas for anisotropic elasticity and polycrystalline averages. Our results provide accurate reference values and an appropriate general methodology for elastic properties of metal halide perovskites.
\end{abstract}

\maketitle
\newpage

\section{Introduction}
Recent research in hybrid organometallic perovskites such as methylammonium lead iodide (MAPI),  has created a new avenue for making low-cost, low-temperature, easy-to-manufacture photovoltaic devices with photoconversion efficiencies (PCE$\sim$27\% \cite{NREL}) comparable to its rival, silicon. On one hand, suitable optoelectronic properties such as direct bandgap \cite{Frohna2018,Umari2014RelativisticGC}, high absorption coefficient \cite{Shirayama2016a}, long diffusion length \cite{Stranks341,Xing344}, and high carrier mobility \cite{Wehrenfennig2014} make perovskites a promising candidate for solar cells, but on the other hand, they suffer from serious degradation problems \cite{niu2015review,conings2015intrinsic,Joshi2016} which hinder their commercialization as solar cells. The presence of oxygen \cite{aristidou2015role} and moisture \cite{niu2015review} rapidly accelerates the degradation in perovskites. However, these factors are not a concern in the vacuum of space, leading to interest for space applications \cite{tu2021perovskite,yang2020potential}. Additional attractive properties are high specific power (23 kW/kg) \cite{kaltenbrunner2015flexible}, low-temperature easy in-situ fabrication in space \cite{mcmillon2022would}, high defect tolerance \cite{steirer2016defect}, excellent radation hardness and self-healing \cite{Miyazawa,Lang}, and lower cost per payload \cite{mcmillon2022would}. Despite these advantages, hybrid perovskite devices have to face some extra challenges in space. For example, they have to go through rapid temperature changes (roughly 300 K) within few hours, withstand high-energy charged particle radiation \cite{lee1977solar}, and face heavy mechanical damage due to space debris travelling at speeds of 10 km/s\cite{moussi2005hypervelocity,delmas2023evaluation}. If the devices are sent up from Earth, they additionally have to endure vibrations during launch and bending in unfolding procedures after launch. \par

Much research has been done to improve the photoconversion efficiency and stability of devices based on MAPI, but little research is available related to its mechanical properties, which are important for making solar cells for Earth or space applications. One reason for the lack of studies may be the complexity of mechanical characterization due to the easily degradable nature of MAPI in ambient conditions \cite{Deretzis,YZhang}. There is no experimental result available so far for cubic and orthorhombic structures reporting mechanical properties, but results are available for the room-temperature tetragonal phase \cite{rathore2021elastic,rakita2015mechanical,liao2021photodegradation,ciric2018mechanical}. Mainly nanoindentation techniques have been used to calculate Young’s modulus at different crystal surfaces, but also atomic force microscopy (AFM) was used to measure the elastic modulus \cite{liao2021photodegradation}. Experiments have report elastic moduli in different crystal directions \cite{rathore2021elastic,sun2015mechanical,rakita2015mechanical,liao2021photodegradation,ciric2018mechanical,rathore2021elastic}, complicating comparisons.
Considering these few available experimental studies on mechanical properties of MAPI, there are significant disagreements between their results (Table \ref{tab:MechProp}).
There are numerous potential reasons for this variation. The probed regions may include grains and grain boundaries \cite{leguy2015dynamics}. Other defects and impurities \cite{Glaser} may be present. Given the material's instability, partial degradation \cite{Deretzis,YZhang} of bulk or surface may occur before measurement. There may be substantial residual strains \cite{Xue}, potentially in inhomogeneous and anisotropic patterns. There also may be coexistence of different crystal structures, or effects of the thin film thickness (e.g. related to the surface or substrate). Moreover there are disagreements with theoretical results. \par

Theoretical studies with density-functional theory (DFT) are available for all three phases but differ widely amongst each other and with experiments \cite{rathore2021elastic, Feng2014, diao2020study,khellaf2021advances,ali2018theoretical,roknuzzaman2018insight} (Table \ref{tab:MechProp}).
These can stem from calculation-related issues such as different pseudopotentials, inadequate relaxation of the structure, inappropriate atomic structures of MAPI, etc., which can be reflected in unusual results, such as negative values for certain elastic constants or inconsistencies in reported lattice parameters across different phases \cite{diao2020study}. In some cases, reported values for mechanical properties like Young’s modulus can differ significantly from other studies \cite{diao2020study}, which could indicate a need to re-evaluate the underlying calculations. In other cases \cite{ali2018theoretical}, there are questions about the appropriateness of formulas used for the calculation of derived quantities such as polycrystalline averages in the Voigt approximation \cite{hill1952elastic}. Recalculating a polycrystalline Young’s modulus using the Hill approximation \cite{hill1952elastic} based on previously reported elastic constants \cite{Feng2014} leads to some differences from the results reported.
Given these considerations, it is crucial thoroughly to check and apply the correct methodologies and formulas when calculating and reporting material properties. This careful approach will help to ensure consistency and reliability in the literature.

In this paper, we have studied the mechanical properties of all three phases of MAPI systematically to determine factors that may cause the disagreement between different theoretical results, as well as disagreement from experimental results, and to obtain clear consistent reference values. We calculated the whole stiffness $C_{ij}$  and compliance matrices $S_{ij}$ for all three phases of MAPI. The paper is organized as follows. In Section \ref{sec:comp_details}, we detail our computational approach. In Section \ref{sec:framework}, we detail our methods for applying strain, extracting elastic moduli, and rotating the tensor to different crystal directions, to ensure clear and reproducible results. In Section \ref{sec:results}, we compare the available experimental and theoretical results. We assess the impact of the method of elastic calculation, exchange-correlation functionals, Van der Waals corrections, pseudopotentials, $k$-point sampling, and structures
and well as appropriate formulas for polycrystalline averages or elastic properties in particular crystal directions. Finally we provide the reference values we have obtained. In Section \ref{sec:conc}, we conclude with our findings on the key factors to ensure consistent calculations of elastic properties.
\section{Methods}
\label{sec:methods}

\subsection{Computational Details}
\label{sec:comp_details}

We have investigated the three phases of MAPI perovskites: pseudo-cubic \cite{talit2020stress}, tetragonal, and orthorhombic \cite{brivio2015lattice}. The orthorhombic structure is exactly symmetric having $D_{2h}$ point group symmetry but the tetragonal structure does not have any symmetry on an atomistic level \cite{talit2023}. There are experiments that report the structure to have space groups I422 and P$4_22_12$, which are subgroups of I4/mcm.\cite{arakcheeva2016ch3nh3pbi3} It is not possible to do a DFT calculation using fractional occupancy of the methylammonium ion orientations within the lattice, and there are two different tetragonal structures that are mostly used in DFT calculations: quasi-I4cm \cite{brivio2015lattice} and quasi-I4/mcm \cite{leppert2016electric}. We have previously found that, for a given calculation approach, changing between these structures leads to differences of around 1-10\% in elastic tensor elements \cite{talit2023}, which are significantly less than some of the differences between calculation approaches. Therefore in this work, we use only the quasi-I4cm structure for our tetragonal MAPI. %
Our calculations use the plane-wave DFT code Quantum ESPRESSO \cite{giannozzi2009quantum, QE_2}. We checked the effect of different exchange-correlation functionals (LDA \cite{perdew1992accurate}, PBE \cite{perdew1996generalized}, and PBEsol \cite{PBESol_PhysRevLett.100.136406}) and used scalar relativistic optimized norm-conserving Vanderbilt (ONCV) pseudopotentials \cite{Hamann} from Pseud$\bar{o}$ D$\bar{o}$j$\bar{o}$ \cite{van2018pseudodojo}) (NC SR ONCVPSP v0.4) with standard accuracy. We have also used a Troullier-Martins \cite{troullier1991efficient} pseudopotential for PBE to compare with the ONCV result. Different half-shifted Monkhorst-Pack $k$-grids are used for Brillouin zone sampling with plane-wave energy cutoff of 
100 Ry for the wave functions.  Variable-cell relaxation is done using convergence thresholds of 0.5 kbar for stress and 1 meV/\AA for atomic forces. We have compared different Van der Waals correction schemes (Grimme-D2 \cite{grimme2006semiempirical}, Grimme-D3 \cite{grimme2010consistent}, Tkatchenko-Scheffler \cite{tkatchenko2009accurate}) as implemented in Quantum ESPRESSO, to handle the interaction between the methylammonium cation and the inorganic lattice. The applied strain range for these calculations is $\pm 1$\%. \par

\subsection{Theoretical Framework}
\label{sec:framework}
\subsubsection{Calculation of Stiffness Constants}

As we have mentioned in the introduction section about large discrepancies  in reported mechanical properties of CH$_3$NH$_3$PbI$_3$ hybrid perovskites, it is important that we clearly mention all the formulas and techniques that we have used to calculate them. A very well known formula to calculate stiffness constants is Hooke's law which states that within elastic limit, stress is proportional to strain and mathematically expressed as $\sigma_{ij}=C_{ijkl}\epsilon_{kl}$ \cite{smith2013engineering}. Here $\sigma_{ij}$ is the stress, $\epsilon_{kl}$ is the strain, and $C_{ijkl}$ is the stiffness constant; $i$, $j$, $k$, and $l$ are the Cartesian directions $x$, $y$, or $z$. $C_{ijkl}$ is a 4$^{th}$ rank tensor and has 81 elements but due to the minor symmetry (index symmetry) %
it reduces to 36 elements and the 4$^{th}$ rank tensor can be mapped to a $6\times6$ matrix $C_{ij}$ using Voigt notation, which can be further reduced to 21 due to the equivalence of the partial derivative ($C_{ijkl}=C_{klij}$). Due to the symmetry in the crystal system, the independent components of the stiffness tensor even further reduce from 21 to 3 for cubic, 9 for orthorhombic and 6 for tetragonal-(I) crystal system \cite{dresselhaus2007group}. \par

To calculate the elastic and mechanical properties for different MAPI structures we need to calculate all the independent components of the stiffness tensor first. There are two different methods by which this can be done: (a) by calculating the stress tensors directly in the plane-wave formalism \cite{Nielsen1985} for different applied strains and using a linear fit, or (b) by calculating the change in energy density of the system
for applied strains and doing a quadratic fit. We have done both and give a comparison of results in Section \ref{sec:results}. We mainly used the second method, change in energy density, as most of the previously published papers have reported their calculated values of elastic properties using this method. We note additionally that the stress tensor is notoriously more sensitive to the plane-wave cutoff than the total energy, and is thus more liable to be unconverged in calculations.\par 

The Taylor expansion of total internal energy $E$ of a crystal under strain can be written as 
\begin{align}\label{eq:taylor_expansion}
E(V, \epsilon) = E(V_0) + \sum\limits_{i}\frac{\partial E}{\partial \epsilon_i} \epsilon_i + \frac{1}{2} \sum\limits_{i,j}\frac{\partial^2 E}{\partial \epsilon_i \partial \epsilon_j} \epsilon_i \epsilon_j + \mathcal{O}(\epsilon^3)
\end{align}
Here $V_0$ and $V$ denote the volume of the unstrained and strained crystal respectively, where $\epsilon_i$ is the strain in the $i^{th}$ direction.  $\partial^2 E / \partial \epsilon_i \partial \epsilon_j$ is also known as the stiffness constant $C_{ij}$ of the crystal. The change in energy density can be expressed as $U=(E(V, \epsilon)-E(V_0)/V_0$. The idea behind calculating $C_{ij}$ coefficients is that we will strain the structure in some specific directions and calculate the change in energy density. From the quadratic fitting of the change in energy density vs $\epsilon$ graph we can calculate the stiffness constant. The values for $i$ and $j$ range between 1 to 6 with indices defined in Voigt notation as:
$1\equiv xx, 2\equiv yy, 3\equiv zz, 4\equiv yz, 5\equiv zx, 6\equiv xy$.

\subsubsection{How to Strain the Crystal: Strain Tensors}
\begin{figure}[ht]
    \centering
    \includegraphics[width=0.4\textwidth]{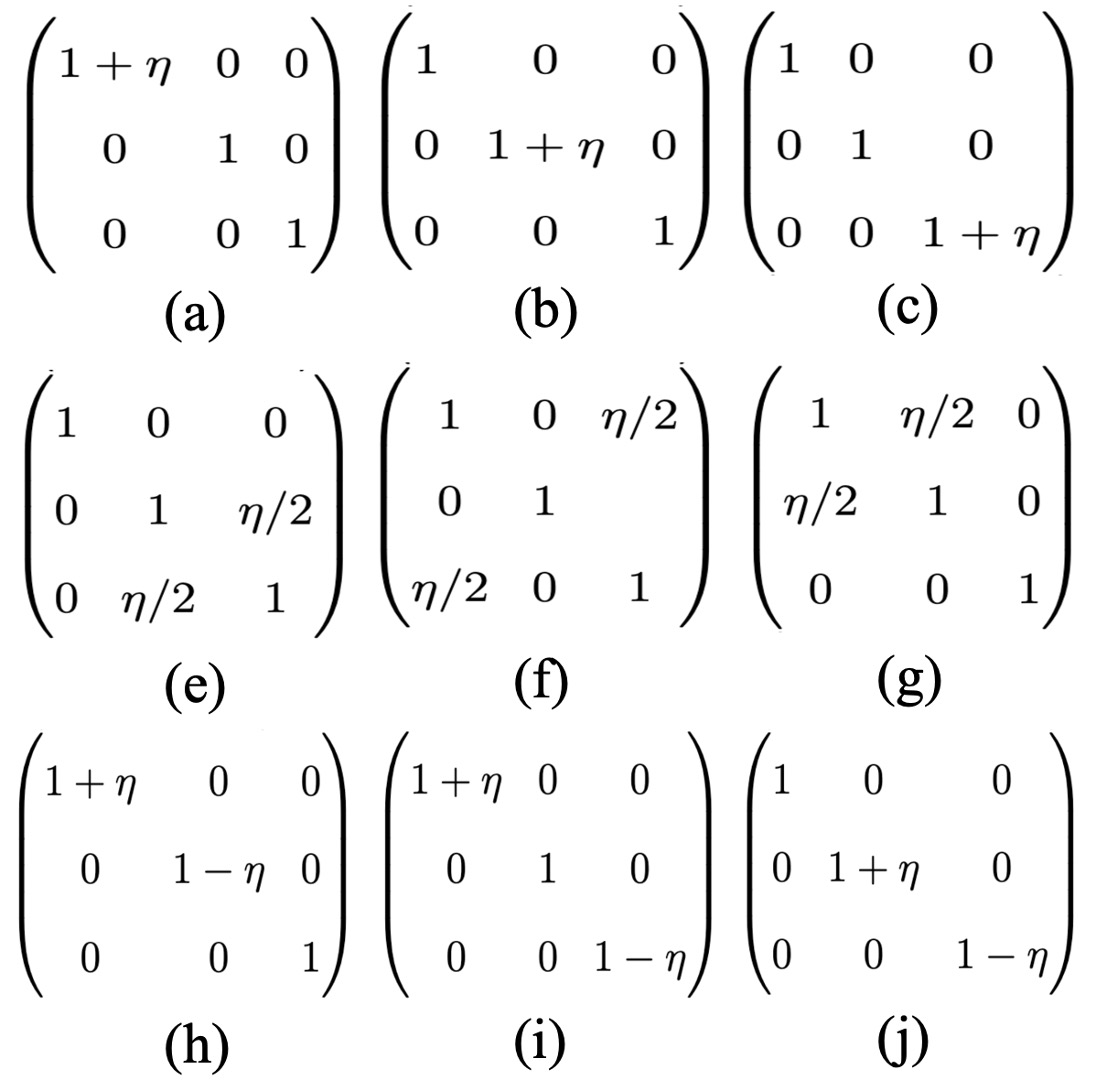}
    \caption{Different strain tensors used to calculate the full stiffness tensor and other mechanical properties. Out of these nine strain tensors a,b and c represent uniaxial strain; e,f, and g represent shear strain; and h, i, and j represent biaxial strain.}
    \label{fig:strain_tensors}
\end{figure}

Every crystal structure has its own lattice parameters (i.e. lattice vectors $\vec{a}=a_x\hat{i}+a_y\hat{j}+a_z\hat{k}$, $\vec{b}=b_x\hat{i}+b_y\hat{j}+b_z\hat{k}$, and $\vec{c}=c_x\hat{i}+c_y\hat{j}+c_z\hat{k}$) and the coordinates of all the atoms in the unit cell can be expressed in terms of the lattice vectors. Defining the coordinates this way, any strain to the lattice vectors automatically applies to the entire system including the atoms. Symmetric strain tensors used for the calculations of different strain directions are given in Fig. (\ref{fig:strain_tensors}). Lattice vectors are transformed by a strain tensor as:
\begin{align}\label{lattice_deformation}
\begin{pmatrix}
\epsilon_{xx} & \epsilon_{xy} & \epsilon_{xz}\\
\epsilon_{yx} & \epsilon_{yy} & \epsilon_{yz}\\
\epsilon_{zx} & \epsilon_{zy} & \epsilon_{zz}
\end{pmatrix} 
\begin{pmatrix}
a_x & b_x & c_x\\
a_y & b_y & c_y\\
a_z & b_z & c_z
\end{pmatrix} =
\begin{pmatrix}
a'_x & b'_x & c'_x\\
a'_y & b'_y & c'_y\\
a'_z & b'_z & c'_z
\end{pmatrix}
\end{align}

\subsubsection{Calculation of different elastic parameters}
After each structure is deformed using the strain tensor as discussed above, we need to relax the ions of these deformed structures keeping the lattice parameter fixed. This process is done using Quantum ESPRESSO. Each deformed, relaxed structure is used to calculate total energy of the system. Once we have it, we can calculate the change in energy density (U) using equation (\ref{eq:taylor_expansion}) which in short can be written as 
\begin{align}\label{energy_density}
    U=\frac{1}{2}\Large\sum_{i=1}^6\sum_{j=1}^6 C_{ij} \epsilon_i \epsilon_j
\end{align}

Using the 9 different strain tensors as mentioned in Fig. (\ref{fig:strain_tensors}) we can calculate 9 independent stiffness constants ($C_{ij}$) using equation (\ref{energy_density}). For $C_{11}$, $C_{22}$, and $C_{33}$ we can use uniaxial strains as shown in Fig. \ref{fig:strain_tensors}(a, b, c) and directly use quadratic fitting to the equation. To calculate $C_{44}$, $C_{55}$, and $C_{66}$ we need to apply shear strain as shown in Fig. \ref{fig:strain_tensors}(e, f, g). We purposefully used $\frac{\eta}{2}$ instead of $\eta$ for the cross terms of $\epsilon$ because there will be 4 terms in the summation and it keeps the final equation simple as $U=\frac{1}{2}C_{ii}\eta^2$. For $C_{12}$, $C_{13}$, and $C_{23}$ we need to apply biaxial strain as given in Fig. \ref{fig:strain_tensors}(h, i, j). Applying these strains will make the energy density equation as $U=\frac{1}{2}(C_{ii}+C_{jj}-2C_{ij})\eta^2$, with $i,j = \{1, 2, 3\}$. Using the quadratic fitting and with previously calculated results of $C_{ii}$ and $C_{jj}$ we can calculate the values of $C_{ij}$. Once we calculate all the independent components of the stiffness tensor for orthorhombic and tetragonal MAPI, we can calculate the compliance tensor which is the inverse of the stiffness matrix. Using the stiffness and compliance matrix we can calculate elastic properties such as Young's modulus ($E$), bulk modulus ($K$), shear modulus ($G$), Poisson's ratio ($\nu$) using  Voigt, Reuss, and Hill approximation\cite{hill1952elastic} which give macroscopic averages for polycrystalline materials.

\subsubsection{Calculation of elastic modulus at different crystallographic directions}
To calculate Young's modulus E at any given direction $\langle hkl \rangle$ \cite{nye1985physical}, we can use the compliance matrix and a relation for orthorhombic structures:
\begin{align}\label{Ehkl}
    \frac{1}{E_{hkl}}= l_1^4 S_{11} + 2 l_1^2 l_2^2 S_{12} + 2 l_1^2 l_3^2 S_{13} + l_2^4 S_{22} + 2 l_2^2 l_3^2 S_{23} \notag \\+ l_3^4 S_{33} + l_1^2 l_3^2 S_{44} + l_1^2 l_3^2 S_{55} + l_1^2 l_2^2 S_{66}
\end{align}
The terms $l_1$, $l_2$, and $l_3$ are the direction cosines 
of the angle between the direction of interest $\langle hkl \rangle$ and the $x$-, $y$-, and $z$-axes.
Some literature has used a form missing the $S_{11}$ term \cite{ravindran1998density}.
For cubic symmetry, equation (\ref{Ehkl}) will reduce to \cite{hopcroft2010young}:
\begin{align}\label{Ehkl_cubic}
    \frac{1}{E_{hkl}}=S_{11} - 2\left(S_{11}-S_{12}-\frac{1}{2}S_{44}\right)(l_1^2l_2^2+l_1^2l_3^2+l_2^2l_3^2)
\end{align}
Formulas (eq. \ref{Ehkl}, \ref{Ehkl_cubic}) for particular crystal systems can be generalized by using the stiffness matrix and transforming it according to using orthogonal transformation matrices:
\begin{align}\label{Cijkl_rotn}
    C^*_{ijkl}=\Omega_{ip}\Omega_{jq}\Omega_{kr}\Omega_{ls}C_{pqrs}
\end{align}
These orthogonal transformation matrices $\Omega$ can be simple rotation matrices. This transformation rule applies to $C_{pqrs}$ which is a $3\times 3\times 3\times 3$ matrix but does not apply directly to our calculated stiffness tensor $C_{\alpha\beta}$ which is a $6\times6$ matrix. We use the transformation law for $C_{\alpha\beta}$ as derived in detail in Ref. \cite{ting1996anisotropic}. 
We constructed a general rotation matrix based on Euler angles:
\begin{widetext}
\begin{align}
    \Omega=
    \begin{pmatrix}
    1 & 0 & 0\\
    0 & \cos\psi & -\sin\psi\\
    0 & \sin\psi & \cos\psi
    \end{pmatrix}
    \begin{pmatrix}
    \cos\beta & 0 & \sin\beta\\
    0 & 1 & 0\\
    -\sin\beta & 0 & \cos\beta
    \end{pmatrix}
    \begin{pmatrix}
    \cos\theta & -\sin\theta & 0\\
    \sin\theta & \cos\theta & 0\\
    0 & 0 & 1
    \end{pmatrix}
\end{align}
To transform the stiffness matrix $C$ due to the rotation $\Omega$, we perform \cite{ting1996anisotropic}:
\begin{align}
    \textbf{C}^* = \textbf{K} \textbf{C} \textbf{K}^T
\end{align}
where 
\begin{align}
    K=\begin{pmatrix}
    K_1 & 2K_2\\
    K_3 & K_4
    \end{pmatrix}
\end{align}
\begin{align}
    K_1=\begin{pmatrix}
    \Omega_{11}^2 & \Omega_{12}^2 &\Omega_{13}^2\\
    \Omega_{21}^2 & \Omega_{22}^2 &\Omega_{23}^2\\
    \Omega_{31}^2 & \Omega_{32}^2 &\Omega_{33}^2
    \end{pmatrix}
\end{align}
\begin{align}
    K_2=\begin{pmatrix}
    \Omega_{12}\Omega_{13} & \Omega_{13}\Omega_{11}  &\Omega_{11}\Omega_{12}\\
    \Omega_{22}\Omega_{23} & \Omega_{23}\Omega_{21}  &\Omega_{21}\Omega_{22}\\
    \Omega_{32}\Omega_{33} & \Omega_{33}\Omega_{31}  &\Omega_{31}\Omega_{32}\\
    \end{pmatrix}
\end{align}
\begin{align}
    K_3=\begin{pmatrix}
    \Omega_{21}\Omega_{31} & \Omega_{22}\Omega_{32}  &\Omega_{23}\Omega_{33}\\
    \Omega_{31}\Omega_{11} & \Omega_{32}\Omega_{12}  &\Omega_{33}\Omega_{13}\\
    \Omega_{11}\Omega_{21} & \Omega_{12}\Omega_{22}  &\Omega_{13}\Omega_{23}\\
    \end{pmatrix}
\end{align}
\begin{align}
    K_4=\begin{pmatrix}
    \Omega_{22}\Omega_{33}+\Omega_{23}\Omega_{32} &  \Omega_{23}\Omega_{31}+\Omega_{21}\Omega_{33}  & \Omega_{21}\Omega_{32}+\Omega_{22}\Omega_{31}\\
    \Omega_{32}\Omega_{13}+\Omega_{33}\Omega_{12} &  \Omega_{33}\Omega_{11}+\Omega_{31}\Omega_{13}  & \Omega_{31}\Omega_{12}+\Omega_{32}\Omega_{11}\\
    \Omega_{12}\Omega_{23}+\Omega_{13}\Omega_{22} &  \Omega_{13}\Omega_{21}+\Omega_{11}\Omega_{23}  & \Omega_{11}\Omega_{22}+\Omega_{12}\Omega_{21}\\
    \end{pmatrix}
\end{align}
\end{widetext}

Once we have the rotated stiffness tensor, we calculate the compliance matrix and thereby calculate the elastic modulus along the rotated crystallographic direction.

\begin{figure*}[ht]
    \centering
    \includegraphics[width=\textwidth]{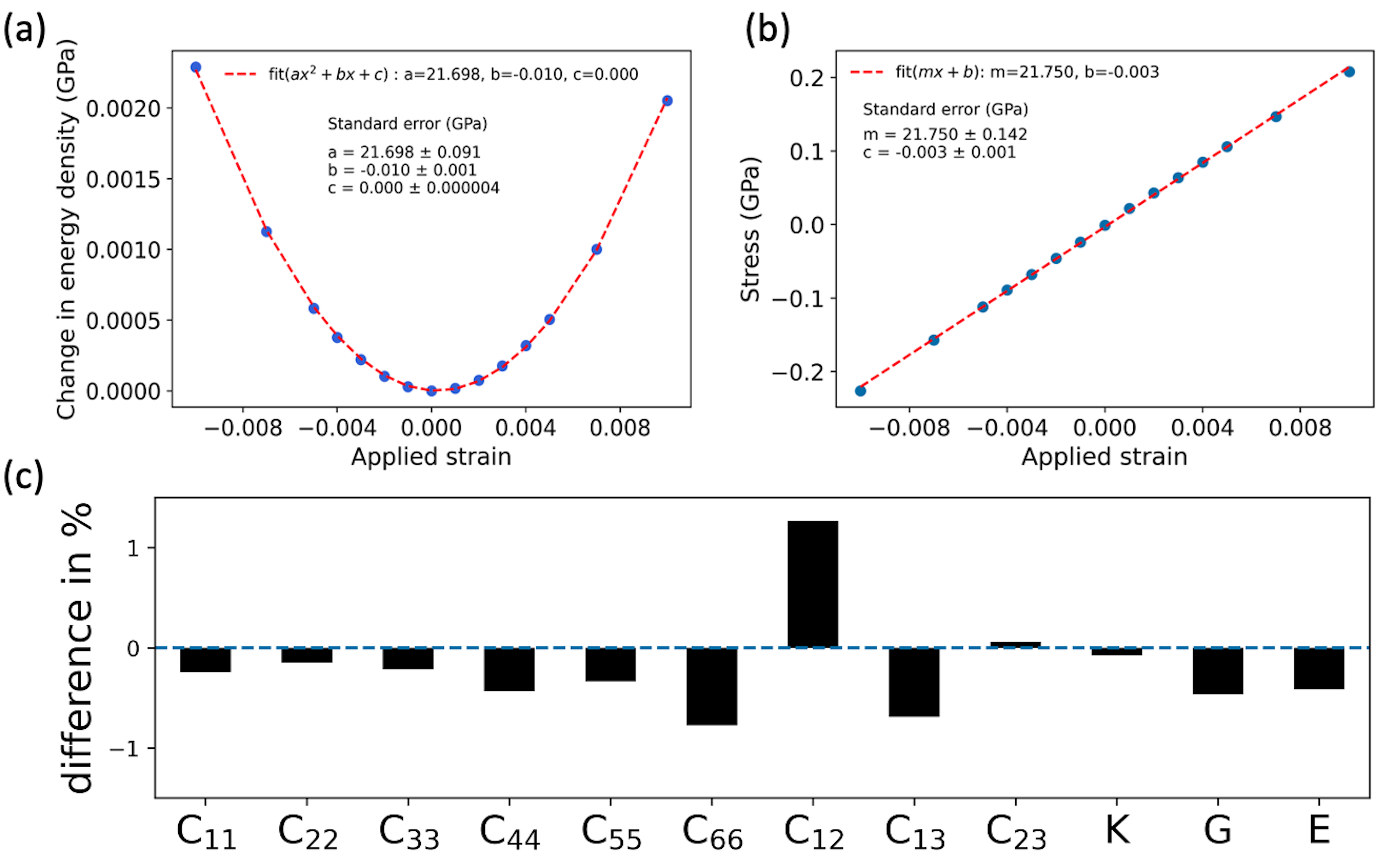}
    \caption{Elastic properties calculated for orthorhombic MAPI structure using two different methodologies. (a) Calculation of $C_{11}$ using energy density method. (b) Calculation of $C_{11}$ using the stress tensor method. (c) Comparison of parameters calculated using two different methods of (a) and (b).}
    \label{fig:method_comp}
\end{figure*}

\begin{table*}[]
\caption{Calculated values of polycrystalline averages (Hill approximation) of bulk modulus $K$, shear modulus $G$, Young's modulus $E$; and values of Young's modulus in crystallographic directions [100], [010], and [001] (all in GPa), comparing previously reported experimental and theoretical results.}
\label{tab:MechProp}
\resizebox{\textwidth}{!}{%
\tiny
\begin{tabular}{rrrrrrrr}
\hline
\multirow{2}{*}{\textbf{Structure}}     & \multicolumn{3}{r}{\textbf{\begin{tabular}[c]{@{}r@{}}Polycrystalline   average\end{tabular}}}                                                                                                                              & \multirow{2}{*}{\textbf{\begin{tabular}[c]{@{}r@{}}$E$\\      {[}100{]}\end{tabular}}} & \multirow{2}{*}{\textbf{\begin{tabular}[c]{@{}r@{}}$E$\\      {[}110{]}\end{tabular}}} & \multirow{2}{*}{\textbf{\begin{tabular}[c]{@{}r@{}}$E$ \\{[}112{]}\end{tabular}}} & \multirow{2}{*}{\textbf{Method}}                                \\
                                        & \textbf{\begin{tabular}[c]{@{}r@{}}$K$\end{tabular}} & \textbf{\begin{tabular}[c]{@{}r@{}}$G$\end{tabular}} & \textbf{\begin{tabular}[c]{@{}r@{}}$E$\end{tabular}} &                                                                                                                  &                                                                                                                  &                                                                                                                  &                                                                 \\ \hline
\multirow{11}{*}{\textbf{Orthorhombic}} & 20.22                                                                         & 7.69                                                                         & 20.48                                                                          & 20.18                                                                                                            & 14.25                                                                                                            & 16.31                                                                                                            & DFT-LDA                                                         \\
                                        & 14.76                                                                         & 6.45                                                                         & 16.88                                                                          & 15.33                                                                                                            & 12.92                                                                                                            & 13.89                                                                                                            & DFT-PBE                                                         \\
                                        & 17.072                                                                        & 7.136                                                                        & 18.79                                                                          & 17.981                                                                                                           & 13.478                                                                                                           & 14.672                                                                                                           & DFT-PBEsol                                                      \\
                                        & 16.70                                                                         & 6.59                                                                         & 17.48                                                                          & 14.96                                                                                                            & 13.55                                                                                                            & 14.54                                                                                                            & DFT-PBE+vdW-GD2                                                 \\
                                        & 16.70                                                                         & 6.87                                                                         & 18.13                                                                          & 14.95                                                                                                            & 13.71                                                                                                            & 14.99                                                                                                            & GD2+dense $k$-grid                                                  \\
                                        & 15.87                                                                         & 6.64                                                                         & 17.48                                                                          & 17.95                                                                                                            & 12.91                                                                                                            & 13.93                                                                                                            & DFT-PBE+vdW-GD3                                                 \\
                                        & 15.41                                                                         & 6.58                                                                         & 17.27                                                                          & 15.92                                                                                                            & 13.33                                                                                                            & 14.42                                                                                                            & DFT-PBE-TM                                                      \\
                                        & 18.10                                                                         & 3.60                                                                         & 15.00                                                                          & -                                                                                                                & -                                                                                                                & -                                                                                                                & DFT-PBE+vdW\cite{Feng2014}           \\
                                        & 40.42                                                                         & 49.10                                                                        & 104.85                                                                         & -                                                                                                                & -                                                                                                                & -                                                                                                                & DFT-PBE-supersoft\cite{diao2020study}          \\
                                        & 16.76                                                                         & 7.32                                                                         & 19.17                                                                          & -                                                                                                                & -                                                                                                                & -                                                                                                                & DFT-PBEsol\cite{ali2018theoretical}            \\
                                        & 15.45                                                                         & 6.59                                                                         & 17.31                                                                          & -                                                                                                                & -                                                                                                                & -                                                                                                                & DFT-PBE\cite{ali2018theoretical}               \\ \hline
\multirow{12}{*}{\textbf{Tetragonal}}   & 20.33                                                                         & 6.84                                                                         & 18.45                                                                          & 14.80                                                                                                            & 31.61                                                                                                            & 15.84                                                                                                            & DFT-LDA                                                         \\
                                        & 13.14                                                                         & 5.33                                                                         & 14.08                                                                          & 10.39                                                                                                            & 23.62                                                                                                            & 14.75                                                                                                            & DFT-PBE                                                         \\
                                        & 17.07                                                                         & 5.76                                                                         & 15.53                                                                          & 14.90                                                                                                            & 24.79                                                                                                            & 13.02                                                                                                            & DFT-PBE+vdW-GD2                                                 \\
                                        & 13.20                                                                         & 7.00                                                                         & 17.70                                                                          & -                                                                                                                & -                                                                                                                & -                                                                                                                & DFT-PBE\cite{rathore2021elastic}               \\
                                        & 12.20                                                                         & 3.70                                                                         & 12.80                                                                          & 9.27                                                                                                             & 22.67                                                                                                            & 7.60                                                                                                             & DFT-PBE+vdW\cite{Feng2014}           \\
                                        & 36.07                                                                         & 6.32                                                                         & -                                                                              & 22.63                                                                                                            & 20.86                                                                                                            & 20.86                                                                                                            & DFT-PBE\cite{diao2020study}                    \\
                                        & 25.61                                                                         & 11.91                                                                        & 30.94                                                                          & 79.91                                                                                                            & 79.89                                                                                                            & 79.83                                                                                                            & DFT-PBE\cite{khellaf2021advances}              \\
                                        & -                                                                             & -                                                                            & -                                                                              & 10.4±0.08                                                                                                        & -                                                                                                                & 10.7±0.05                                                                                                        & Nanoindentation\cite{sun2015mechanical}        \\
                                        & 13.90                                                                         & 5.40                                                                         & -                                                                              & 14.3±1.7                                                                                                         & -                                                                                                                & 14.0±2.0                                                                                                         & Nanoindentation\cite{rakita2015mechanical}     \\
                                        & -                                                                             & -                                                                            & -                                                                              & 15.14                                                                                                            & -                                                                                                                & -                                                                                                                & SPM/AFM\cite{liao2021photodegradation}         \\
                                        & -                                                                             & -                                                                            & 16.5±1.973                                                                     & -                                                                                                                & -                                                                                                                & -                                                                                                                & Nanoindentation\cite{rathore2021elastic}                           \\
                                        & -                                                                             & -                                                                            & -                                                                              & 20±1.5                                                                                                           & -                                                                                                                & -                                                                                                                & Nanoindentation\cite{ciric2018mechanical}      \\ \hline                                  
\multirow{4}{*}{\textbf{Cubic}}         & 18.90                                                                         & 6.04                                                                         & 16.26                                                                          & 33.74                                                                                                            & 12.65                                                                                                            & 9.94                                                                                                             & DFT-LDA\cite{talit2020stress}                                                         \\
                                        & 16.40                                                                         & 16.40                                                                        & 22.20                                                                          & -                                                                                                                & -                                                                                                                & -                                                                                                                & DFT-PBE+vdW\cite{Feng2014}           \\
                                        & 54.08                                                                         & 2.93                                                                         & -                                                                              & -                                                                                                                & -                                                                                                                & -                                                                                                                & DFT-PBE-supersoft\cite{diao2020study}          \\
                                        & 14.70                                                                         & 7.00                                                                         & 18.10                                                                          & -                                                                                                                & -                                                                                                                & -                                                                                                                & DFT-PBE-ultrasoft\cite{roknuzzaman2018insight} \\ \hline
\end{tabular}%
}
\end{table*}

\section{Results and Discussion}
\label{sec:results}

\begin{figure}[ht!]
    \centering
    \includegraphics[width=0.6\textwidth]{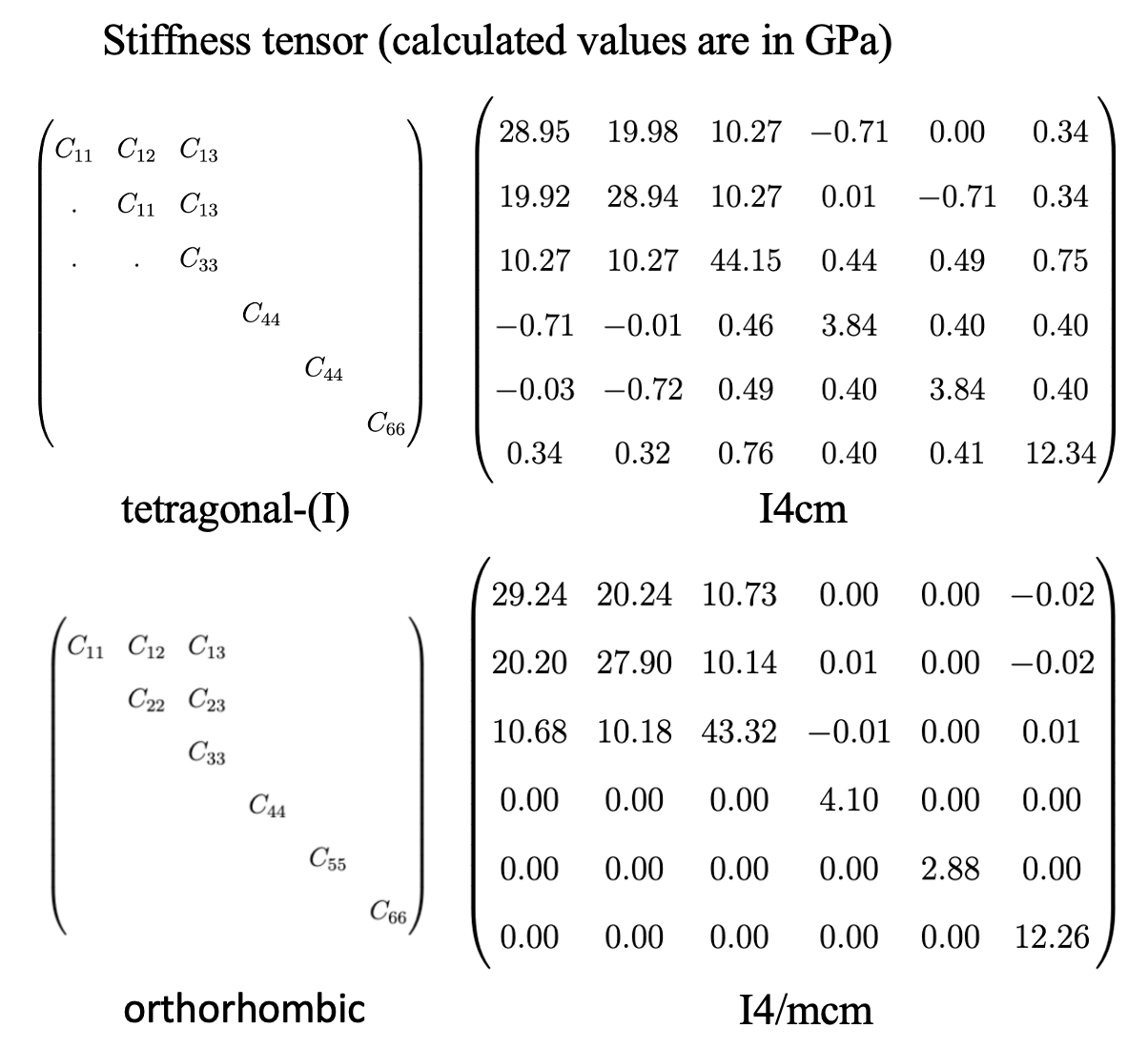}
    \caption{Stiffness matrix calculated for quasi-I4cm and  quasi-I4/mcm structures and compared with the tetragonal-(I) \cite{mouhat2014necessary} and orthorhombic symmetry using LDA.}
    \label{fig:stiffness_tensor}
\end{figure}

To investigate discrepancies in reports about calculated and measured mechanical properties, we began by checking the difference in elastic properties calculated using two different methods: (a) by calculating the energy density and quadratic fit, and (b) by calculating the stress tensor and linear fit. The PBE exchange-correlation functional with ONCV pseudopotential is used for this comparison. The Brillouin zone is sampled using $5\times 4\times 5$ half-shifted Monkhorst-Pack grid with energy cut-off of 100 Ry for the wave functions.
The elastic moduli such as Young's modulus, bulk modulus, and shear modulus using Voigt-Reuss-Hill approximations for orthorhombic MAPI are shown in Fig. \ref{fig:method_comp}. The calculation of stiffness constant $C_{11}$ is given in Fig. \ref{fig:method_comp}(a,b). The difference for all the 9 independent stiffness constant is given in Fig. \ref{fig:method_comp}(c)). It can be seen that the calculated elastic parameters directly from the stress tensor are little higher for most of the cases but the differences between two methods are much smaller than the typical experimental error bar ($\pm 2\%- \pm 10\%$) and the level of discrepancy in theory. So, this difference in calculation methodology for elastic properties calculation does not matter much. For the rest of our calculation, we have used the energy density method, to be more comparable to previously published results. Our calculation of the full stiffness tensors for two tetragonal structures \cite{talit2023} is shown in Fig. \ref{fig:stiffness_tensor}.

\begin{figure}[ht]
    \centering
    \includegraphics[width=0.6\textwidth]{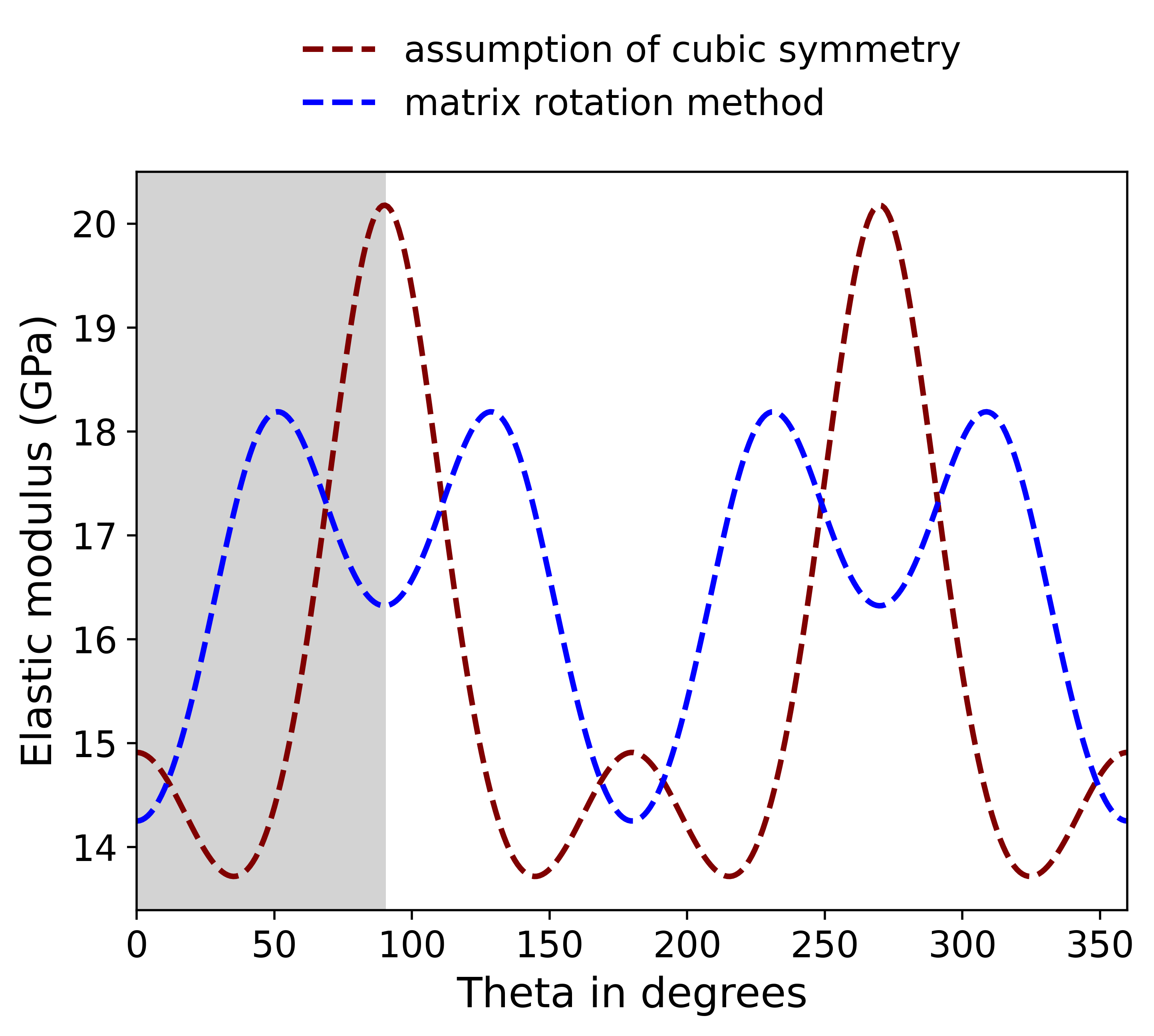}
    \caption{Calculation of elastic modulus as a function of angle from [100], in the (110) plane, for orthorhombic MAPI. The general stiffness matrix rotation method is compared to results from an inappropriate assumption of cubic symmetry.}
    \label{fig:emod_profile}
\end{figure}

\begin{figure}[ht]
    \centering
    \includegraphics[width=\textwidth]{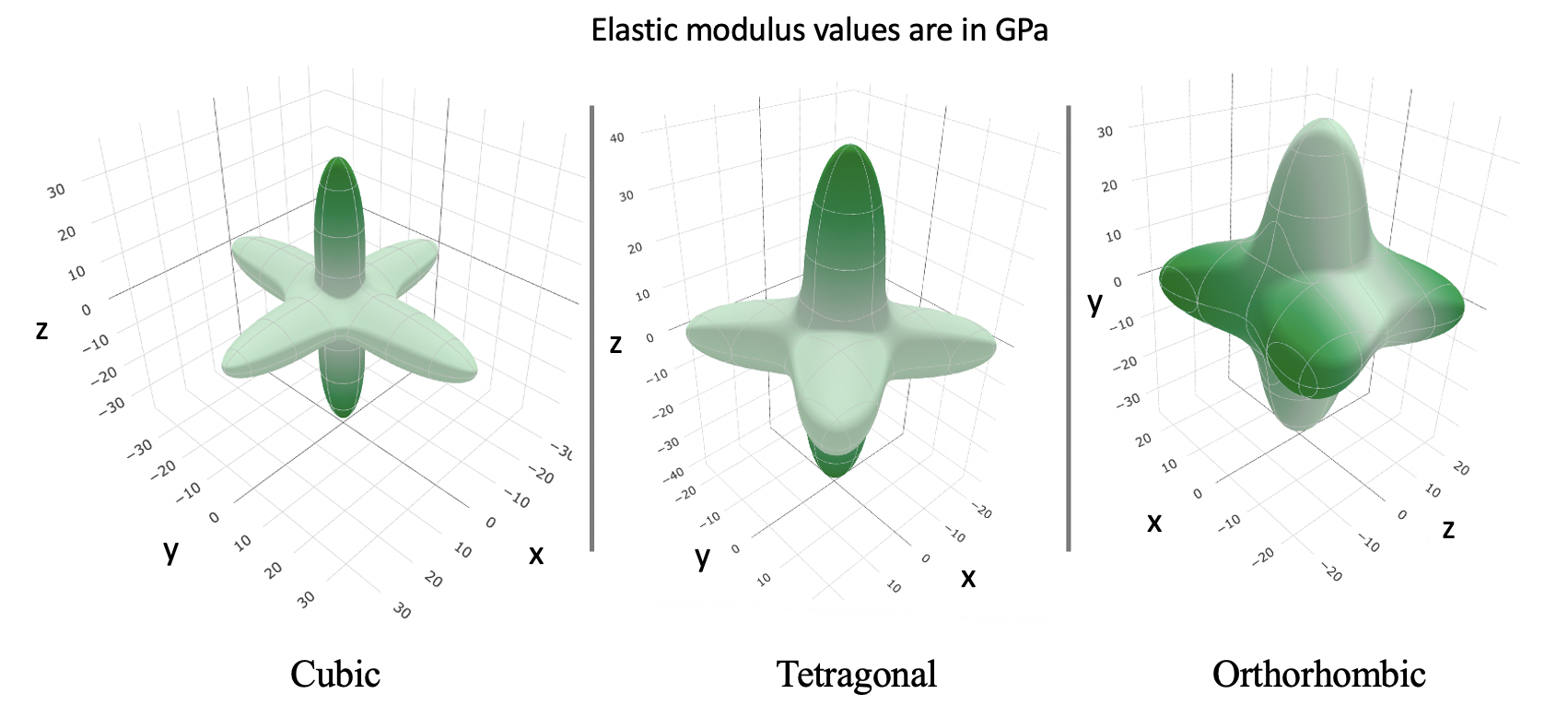}
    \caption{3D profile of the elastic modulus for cubic, tetragonal, and orthorhombic MAPI, visualized using ELATE \cite{gaillac2016elate}.}
    \label{fig:elate_pic}
\end{figure}

We have mentioned in our theoretical framework how to calculate Young's modulus at a particular crystallographic direction in two different ways. The reason we choose to transform the stiffness matrix first and then calculate the elastic modulus on the rotated stiffness matrix is that we do not need to know the exact formula for any particular crystal system. Use of formulae for an incorrectly assumed cubic symmetry, i.e. simplification of equation (\ref{Ehkl}) to equation (\ref{Ehkl_cubic}), will cause different results. Even the high-temperature MAPI is not exactly cubic but pseudo-cubic \cite{talit2020stress}. We can get the compliance matrix from the transformed stiffness matrix and calculate the elastic modulus just using general formula ($E_{11}=1/S_{11}$) where $E_{11}$ is the elastic modulus along the rotated $x$ axis of the crystal system. This general method works for any system. It is even more helpful when we have an approximately symmetric structure rather than an exact one \cite{talit2023}. Using the general method, we also lower the chance of using  formulae based on the incorrect assumption of symmetry as in some works. We have calculated the elastic modulus profile in the (110) plane of orthorhombic MAPI using the general method and the one assuming cubic symmetry (Eq. \ref{Ehkl}), shown in Fig. \ref{fig:emod_profile}. The incorrect assumption of symmetry may lead to substantial difference. One more thing to be noticed in this figure is that it is symmetric in 4 quadrants of the (110) plane and 0-90$^0$ is enough to understand the elastic modulus profile in the plane. For the rest of our results, we have shown variation of elastic modulus data for this range only. Once we have the stiffness tensor for a system, we can draw a 3D profile of the elastic modulus using ELATE \cite{gaillac2016elate}. It can be seen that the Young’s modulus is highest parallel to the Pb-I-Pb bond and highest along the largest lattice parameter in each case. Tetragonal MAPI has the highest value of Young’s modulus among the three phases (Fig. \ref{fig:elate_pic}). 

\begin{figure*}[ht]
    \centering
    \includegraphics[width=0.9\textwidth]{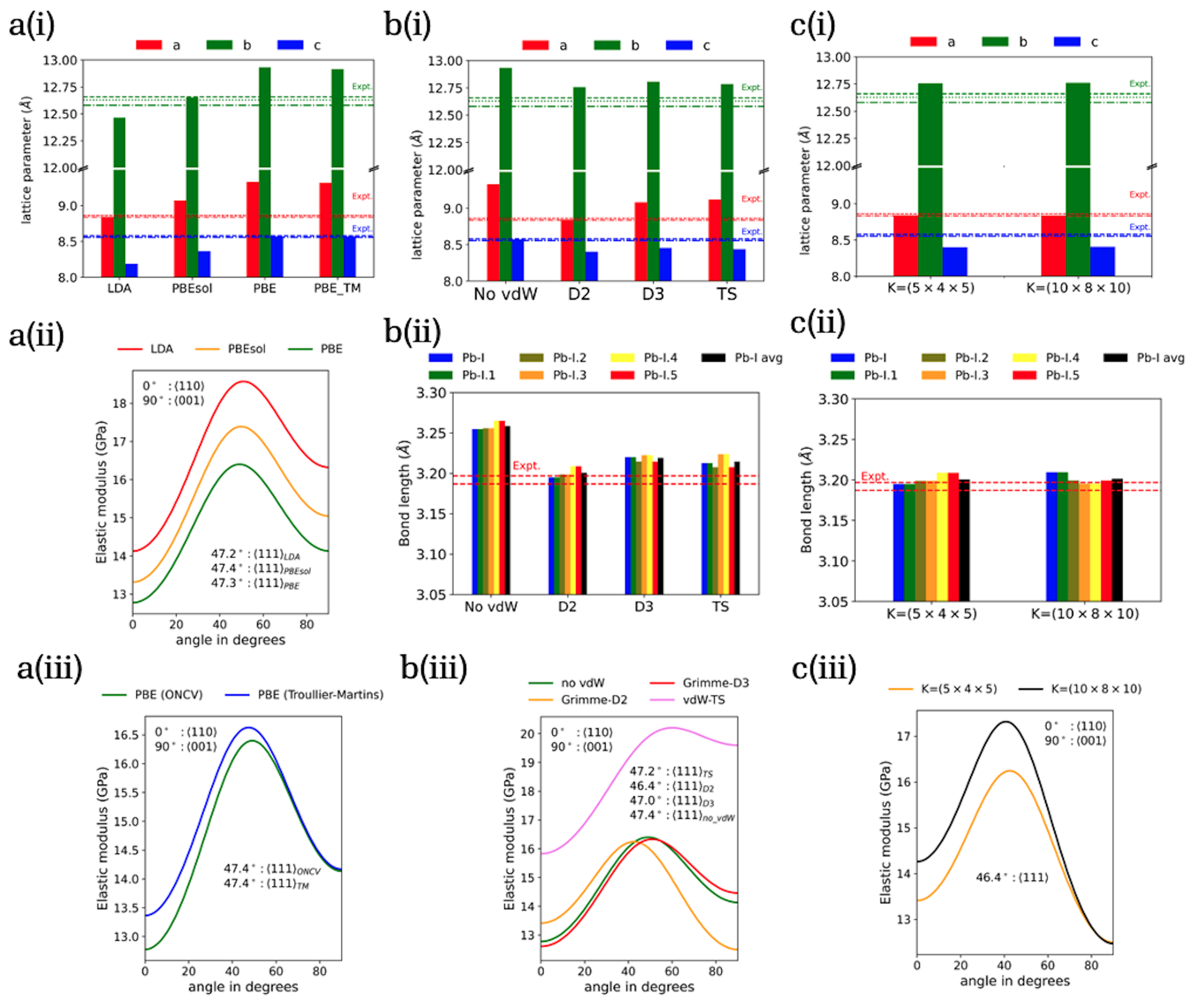}
    \caption{Effect on lattice parameters, bond lengths, and elastic moduli of orthorhombic MAPI, for (a(i,ii)) different exchange-correlation functionals, (a(iii)) pseudopotentials, (b) Van der Waals correction schemes, and (c) $k$-point sampling. Experimental results from \cite{poglitsch1987dynamic,baikie2013synthesis, PND}  }.
    \label{fig:comp_all}
\end{figure*}

\begin{figure*}[ht]
    \centering

    \begin{subfigure}[b]{0.8\textwidth}
        \centering
        \includegraphics[width=\textwidth]{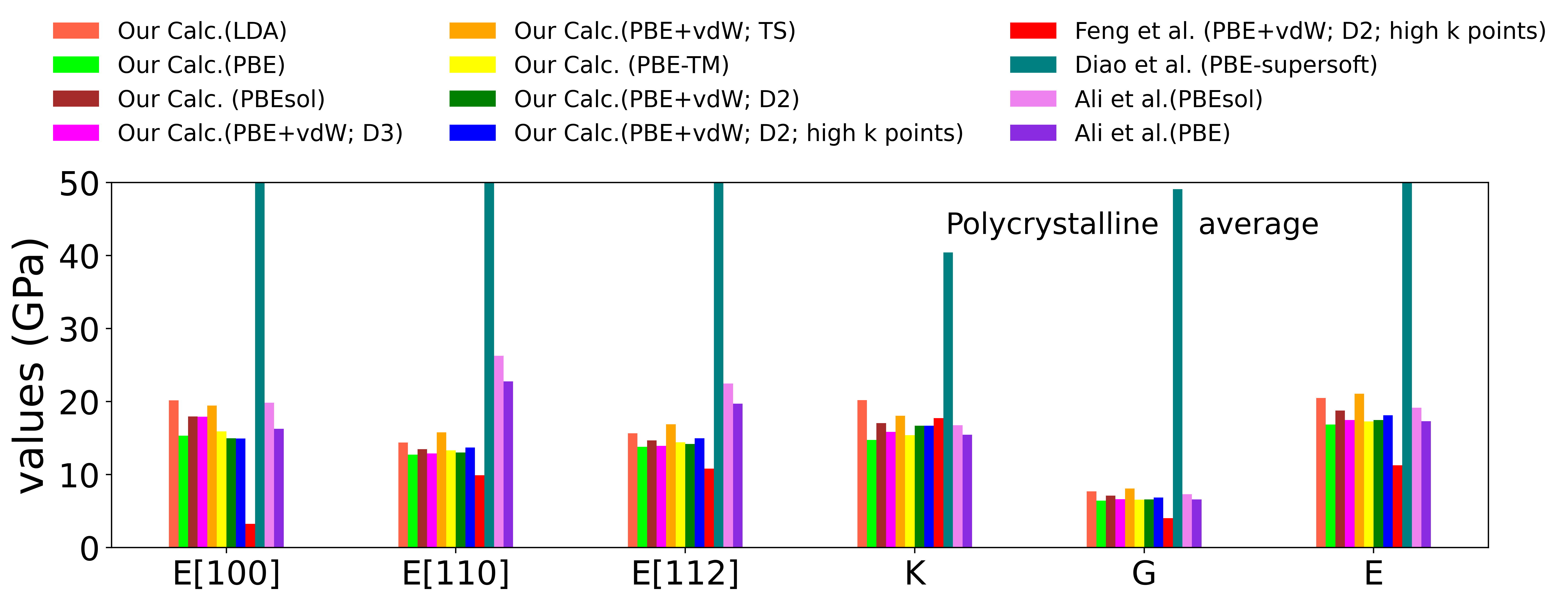} %
        \caption{Orthorhombic MAPI}
        \label{fig:OMAPI}
    \end{subfigure}

    \vspace{0.5cm} %

    \begin{subfigure}[b]{0.8\textwidth}
        \centering
        \includegraphics[width=\textwidth]{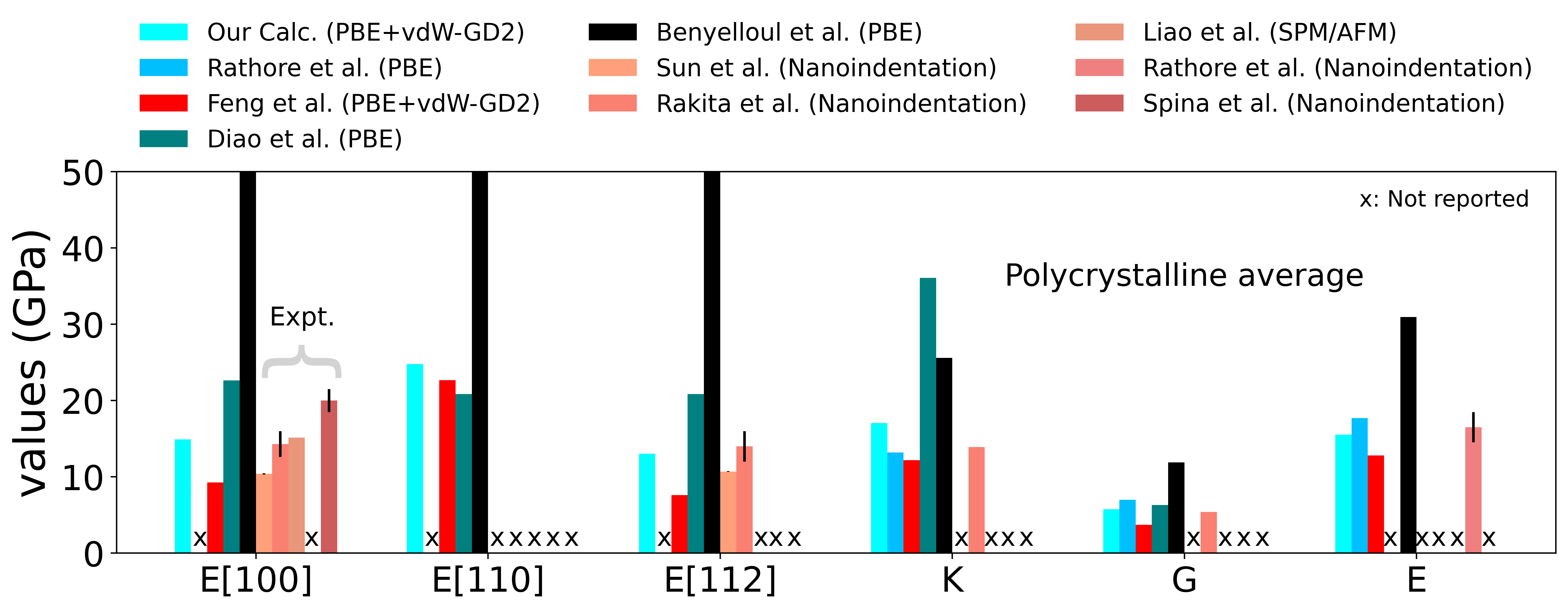} %
        \caption{Tetragonal MAPI}
        \label{fig:TMAPI}
    \end{subfigure}

    \caption{Comparison of our calculated elastic properties with the published theoretical results: Feng et al.\ \cite{Feng2014}, Diao et al.\ \cite{diao2020study}, Ali et al.\ \cite{ali2018theoretical}, Benyelloul et al.\ \cite{khellaf2021advances}; and experimental results: Sun et al.\ \cite{sun2015mechanical}, Rakita et al.\ \cite{rakita2015mechanical}, Liao et al.\ \cite{liao2021photodegradation}, Rathore et al.\ \cite{rathore2021elastic}, Spina et al\ \cite{ciric2018mechanical}.}
    \label{fig:comp_barchart}
\end{figure*}

To investigate further different aspects of calculating elastic parameters, we choose orthorhombic MAPI as it has exact symmetry. The outcome is similar for cubic and tetragonal structures. The effect of different exchange-correlation functionals and pseudopotentials on the orthorhombic MAPI structure and on its elastic modulus is shown in Fig. \ref{fig:comp_all}a(ii). We can see that lattice parameters and Pb-I bond lengths are underestimated by LDA and overestimated by PBE, and the corresponding elastic modulus profile in (110) plane (Fig. \ref{fig:comp_all} a(ii)) shows the opposite. PBEsol stays in-between in both the cases. The exchange-correlation functionals play a major role in calculating the correct structure. Bokdam \textit{et al.} have done a detailed study of MAPI based on the many-body energy in the random-phase approximation (RPA) and find that the strongly constrained and appropriately normed (SCAN) functional \cite{SCAN} is the best for calculating correct structures for hybrid perovskites \cite{bokdam2017assessing}. A future interesting question to study is how SCAN functionals perform for the elastic properties of hybrid perovskite.

We also looked at effect of different pseudopotentials. We compared the results between ONCV (from pseudodojo \cite{van2018pseudodojo}) and Troullier-Martin (from Quantum Espresso \cite{giannozzi2009quantum}) pseudopotentials. The results are similar, with only slight differences in the elastic properties which can be understood from slight difference in their calculated structures (Fig. \ref{fig:comp_all}a(i,iii)). Including Van der Waals corrections to the exchange-correlation functional improve the structural parameters. It affects the elastic modulus mostly along $\langle 110 \rangle$ and $\langle 001 \rangle$, as shown in Fig. \ref{fig:comp_all}b(iii). The MA$^+$ ion is aligned in the (110) plane in the orthorhombic structure, which makes all the H atoms out of this plane and along $\langle 110 \rangle$. Among different available schemes of Van der Waals correction in Quantum ESPRESSO, Tkatchenko-Scheffler (TS) calculates the Young’s modulus higher than average values. Grimme-D3 does not improve the elastic modulus much from PBE alone. Grimme-D2 affects results most along the largest direction $\langle 001 \rangle]$ of the crystal. In future work, other schemes can be studied such as the many-body dispersion (MBD) correction \cite{MBD}, which is by construction more accurate than pairwise Van der Waals interactions, or the exchange-hole dipole method (XDM) \cite{XDM}.

We find that $k$-point sampling plays a crucial role in calculating elastic modulus at particular directions (Fig. \ref{fig:comp_all}c). We have seen that high k-point sampling ($10\times 8\times 10$) makes a difference around 1 GPa in elastic modulus along $[110]$ and $[111]$ directions and no change along $[010]$. Although high $k$-point sampling has significant effect in calculating elastic modulus at a particular direction, it does not have much effect on lattice parameters calculated using ($5\times 4\times 5$) $k$-point grid which is sufficient for converging the total energy by usual criteria. High $k$-point sampling also does not affect the polycrystalline averages of the elastic modulus (Fig. \ref{fig:comp_barchart}a). One hand, we know that PBE with Van der Waals correction (Grimme-D2) gives good result, but we need to have high $k$-point sampling along with a high energy cutoff which needs more computation time. Use of ultrasoft pseudopotentials \cite{ultrasoft}, allowing lower planewave energy cutoffs, could potentially save some computation time without sacrificing the accuracy; these appear to have been used under the term ``supersoft'' in Ref. \cite{diao2020study}. In addition to that, some different methodology such as projector augmented wave (PAW) \cite{Blochl} and all-electron methods would be useful to check on how they affect the mechanical properties of hybrid perovskites. The spin-orbit coupling does not affect the structural parameters significantly near equilibrium \cite{brivio2015lattice} and hence is expected not to make much difference in the calculation of elastic properties.

Our calculated results for both orthorhombic and tetragonal MAPI structures are summarized in Fig. \ref{fig:comp_barchart}. We can see that some of the reported elastic parameters are signficantly higher than the rest of the values, which may very well happen for inaccurate structures (due to problems in the starting point or in structural optimization). It can be seen that LDA overestimates and PBE underestimates all the elastic properties. PBEsol gives better values for elastic modulus but PBE with Grimme-D2 Van der Waals correction gives the best result. With high $k$-points elastic modulus improves in certain directions but it does not change very much for the poly-crystalline averages. We can also see that no experimental results are available in case of orthorhombic structure to validate the result. Since PBE with Grimme-D2 vdW correction with high k-point works best, we used the same for tetragonal structure and validate our results with the experimentally available results (Fig. \ref{fig:comp_barchart}(b)). Our results, with nominally similar parameters to the commonly cited work of Feng \cite{Feng2014}, provided better agreement with the experimental measurements.

\section{Conclusion}
\label{sec:conc}
Knowledge of the elastic properties of hybrid perovskites is important for a variety of applications at large scale. Surprisingly, after more than a decade of hybrid perovskite research for solar cell applications, we have very few reports available that studied the elastic properties of MAPI, and there are substantial discrepancies among experimental and computational results. In this study we have done a detailed analysis of different aspects of the theoretical calculations with DFT to understand the root causes of discrepancies among computational results. While the use of the energy density or stress method and different pseudopotentials makes little difference, the correct use of symmetry in handling the anisotropic stiffness matrix is important.  The exchange-correlation functional is quite important, especially as it affects the structural properties and their change under strain. The Van der Waals correction to the exchange-correlation functional plays a key role for orthorhombic and tetragonal structure. PBE with Grimme-D2 Van der Waals correction with high $k$-point sampling works best for calculating elastic properties. One thing to note here is that use of high $k$-points does not affect the polycrystalline averages that much but takes a lot of computational time, so with a trade-off between computation time and accuracy the high $k$-points need only be used for study of the anisotropy of the elastic modulus. Other methodological approaches of interest for future testing include the SCAN functional, MBD and XDM Van der Waals corrections, spin-orbit coupling, and ultrasoft and PAW pseudopotentials.
Our detailed explanation of the calculation methodology, supported by explicit equations, helps clarify issues that have caused confusion in the literature and can aid in the calculation of elastic properties for complex hybrid perovskite materials, especially in the absence of exact symmetries.

\begin{acknowledgments}
This work was supported by the Merced nAnomaterials Center for Energy and Sensing (MACES), a NASA-funded research and education center, under award NNH18ZHA008CMIROG6R. This work used computational resources from the Multi-Environment Computer for Exploration and Discovery (MERCED) cluster at UC Merced, funded by National Science Foundation Grant No. ACI-1429783, and the National Energy Research Scientific Computing Center (NERSC), a U.S. Department of Energy Office of Science User Facility operated under Contract No. DE-AC02-05CH11231.
\end{acknowledgments}

\bibliography{main} 

\end{document}